\documentclass[pre,superscriptaddress,showpacs,eqsecnum,twocolumn]{revtex4}
\usepackage{newlfont}
\usepackage{amssymb}
\usepackage{amsfonts}
\usepackage{amsmath}
\usepackage{graphicx}
\usepackage{bm}
\begin{document}
\title{Classical fields approximation for cold weakly interacting bosons without free parameters}
\author{{\L}ukasz Zawitkowski}
\affiliation{Center for Theoretical Physics, Polish Academy of Sciences, Aleja Lotnik{\'o}w 32/44, 02-668 Warszawa, Poland}
\author{Miros{\l}aw Brewczyk}
\affiliation{Uniwersytet w Bia{\l}ymstoku,ul. Lipowa 41, 15-424 Bia{\l}ystok, Poland}
\author{Mariusz Gajda}
\affiliation{Institute of Physics, Polish Academy of Sciences, Aleja Lotnik{\'o}w 32/44, 02-668 Warszawa, Poland}
\author{Kazimierz Rz{\c a}{\.z}ewski}
\affiliation{Center for Theoretical Physics, Polish Academy of Sciences, Aleja Lotnik{\'o}w 32/44, 02-668 Warszawa, Poland}
\affiliation{Faculty of Mathematics and Natural Sciences, Cardinal Stefan Wyszy\'nski University, Aleja Lotnik{\'o}w 32/44, 02-668 Warszawa, Poland}

\date{\today}

\begin{abstract}
Classical fields approximation to cold weakly interacting bosons allows
for a unified treatment of condensed and uncondensed parts of the
system. Until now, however, the quantitative predictions were limited by
a dependence of the results on a grid chosen for numerical implementation
of the method. In this paper we propose replacing this unphysical
ambiguity by an additional postulate: the temperature of the gas at
thermal equilibrium should be the same as that of an ideal Bose gas with
the same fraction of condensed atoms. As it turns-out, with this
additional assumption, nearly all atoms are within the classical fields,
thus the method applies to the whole system.
\end{abstract}

\maketitle

\section{Introduction}

The experimental realization of a Bose-Einstein condensate in dilute alkali atomic gases \cite{BEC1,BEC2,BEC3} has
opened a novel possibilities to explore the behavior of large quantum systems.
A study of an interplay between classical and quantum regimes is now available, since 
the experiments are carried out at very low, yet non-zero temperatures. This is an important issue,
because many of these results depend on the temperature.

The atomic BEC is a many-body quantum interacting system, thus
it is not easy to build a usable dynamical theory, that takes thermal effects into consideration.
One group of models addresses this task in a two-gas approach, by explicitly dividing the system into the condensate and the thermal cloud 
\cite{two-gas,two-gas2,two-gaszmag2,two-gas3popov}.
Both subsystems evolve with different equations and they influence each other by a mean-field interaction.
Such an arbitrary separation does not however allow for a deeper insight into the dynamics of the BEC.

The second approach, explored in this paper, is the classical fields approximation (CFA, or its canonical version known as the Wigner function method)
\cite{Castin,Davis,CFAinni,Goral1}. It is a mean-field approximation of the quantum field theory of
interacting bosons, in which one identifies the "classical modes" and evolves them using 
the Gross-Pitaevskii equation (GPE) on a grid. The need to choose the finite number of modes for simulation introduces a cut-off parameter into this 
method.
The approximation does not allow to describe quantum correlations, but it lets one observe the emergence of the
BEC from the pure Hamiltonian dynamics. This is remarkable as condensed and non-condensed atoms are described in the
same way in the classical fields approximation, and the interaction and the observation process allows to distinguish between the two phases.

In this paper we present a version of the classical fields approximation in a 3D box potential without free parameters. 
This goal is achieved by introducing a transformation from numerical control parameters: the energy per particle $E$, an 
interaction strength $g$ and a grid size $n_{grid}$, to physical control parameters: the number of particles 
$N$, the temperature $T$ and the scattering length $a$. Such transformation is not uniquely 
defined; in fact, it constitutes the major difficulty in interpretation of the classical fields approximation. 
The main problem in applying the CFA has been to properly tune the grid size $n_{grid}$ to the temperature and the number of particles.
In previous applications authors used different approaches.
One of them is to fix the population of the highest momentum mode to some arbitrary value and calculate number of particles
a posteriori \cite{Peter,Goral1,Goral_opt,wiry}.
The other one \cite{Davis,DavisBurnett} is to set the numerical grid and the number of atoms on the grid, while treating the classical
fields as just a small subset of a bigger system.

In this paper we suggest yet another method of assigning the physical parameters to a given numerical simulation. The main idea is to set the
temperature of the system to the value of an ideal gas with the same condensate fraction. This is of course an approximation, since it is known that
the critical temperature depends on the interaction strength. However, in the weakly interacting regime a shift of the critical temperature is very
small.

The paper is organized as follows. In Section \ref{The Method} we introduce the classical fields approximation. 
We identify numerical control parameters that determine the properties of the equilibrium of the system. 
Next we review eigenmodes of the system - the generalized Bogoliubov quasiparticles.

In Section \ref{A transition to BEC} we present an analysis of the thermal equilibrium of the system. We observe a phase transition as 
an abrupt change of properties of the system and show that the equipartition of energies in eigenmodes of the system occurs in a condensed phase only. 
We determine the dimensionless temperature per particle $\tau$, which is proportional to $T/N$ \cite{DavisBurnett,Peter}. We also explain a pitfall that one
encounters when trying to define the method via setting unphysical parameters: the grid size or the population cut-off.

In Section \ref{Disposing of cut-off} we compare the scaling properties of the temperature of a system obtained from equipartition of energies with the
temperature of the ideal Bose gas. From this we derive a formula for the number of particles and obtain a set of physical control parameters
($N$, $T$, $a$, $L$) without referring to the grid size. Next, we perform an approximate check of consistency by
comparing the numerical kinetic energy per particle with a result for an ideal gas and find an agreement within $40\%$. Finally, we confront the 
calculated population cut-off with assumptions of the classical fields approximation.

In Section \ref{Conclusions} we present some conclusions.

\section{The Method. Normal modes and their energies}\label{The Method}
Let us consider a gas of N identical bosons trapped within a 3D box potential of length L with periodic 
boundary conditions. We assume that atoms interact via a contact potential 
$V(r-r^\prime)=4\pi\hbar^2a\delta(r-r^\prime)/m$, where $a$ is the s-wave scattering length, known to be adequate 
at low temperatures.

The second-quantized Hamiltonian reads:
\begin{equation}\label{Hamiltonian}
  H=\int_{L^3} d^3x \ (\hat{\Psi}^\dagger \frac{\hat{p}^2}{2m} \hat{\Psi})+\frac{2\pi^2\hbar^2a}{m} 
\int_{L^3} d^3x \ (\hat{\Psi}^\dagger \hat{\Psi}^\dagger \hat{\Psi}\hat{\Psi)},
\end{equation}
where $\hat{\Psi}$ is a bosonic field operator satisfying equal time bosonic commutation relation 
$[\hat{\Psi}(r,t), \hat{\Psi}^{\dagger}(r^\prime,t)]=\delta(r-r^\prime)$.
The Heisenberg equation for $\hat{\Psi}$ resulting from this Hamiltonian is of the form:
\begin{equation}\label{Heisenberg}
  i\hbar\partial_t \hat{\Psi}=-\frac{\hbar^2\Delta}{2m}\ \hat{\Psi}+\frac{4\pi\hbar^2a}{m} \hat{\Psi}^\dagger \hat{\Psi}\ \hat{\Psi}.\\
\end{equation}

The symmetry of the box with periodic boundary conditions sets a natural basis of plane waves with 
quantized momentum $\textbf{p}=2\pi \hbar \textbf{k}/L$, thus it is convenient to expand the field operator:
\begin{equation}\label{Decomposition}
  \hat{\Psi}=\frac{1}{\sqrt{L^3}} \sum_{\textbf{k}} e^{-2\pi i \textbf{kr}/L} \hat{a}_{\textbf{k}}(t).
\end{equation}
Annihilation operators $\hat{a}_{\textbf{k}}$ destroy particle in mode $\textbf(k)$ and satisfy a commutation relation
$[\hat{a}_{\textbf{k}},\ \hat{a}^{\dagger}_{\textbf{k}^\prime}]=1$.

Using the above decomposition in Eq. (\ref{Heisenberg}) we get a set of nonlinear operator 
equations for annihilation operators of plane wave modes: 
\begin{equation}\label{PlaneHeisenberg}
  \partial_t \hat{a}_{\textbf{k}}(t)=-i\frac{2\pi^2\hbar}{mL^2}\ k^2\ \hat{a}_{\textbf{k}}(t)-i\frac{4\pi\hbar a}{mL^3} 
\sum_{\textbf{q}_1,\textbf{q}_2} 
  \hat{a^\dagger}_{\textbf{q}_1}\hat{a}_{\textbf{q}_2}\hat{a}_{\textbf{k}+\textbf{q}_1-\textbf{q}_2}
\end{equation}

A full solution of these equations is not known. However we can simplify this problem by applying 
an approximation which is an extension of the Bogoliubov approach. We extract from the field operator 
all modes that are occupied by sufficiently large number of atoms, such that one can justify neglecting their quantum nature, 
and replace their annihilation operators with c-numbers: 
\begin{equation}\label{CFA}
  \hat{a}_{\textbf{k}} \rightarrow \sqrt{N} \alpha_{\textbf{k}}
\end{equation}

This approximation has been extensively used in description of a multimode laser light in quantum optics; 
here it corresponds to an assumption that these "classical modes" are coherent. The square of the 
modulus of the amplitude $|\alpha_{\textbf{k}}|^2$ gives the fraction of atoms that occupy the mode $\textbf{k}$.

Our set of classical modes corresponds to momenta forming a 3D cubic grid with $n_g$ points in each 
direction. The size of this lattice is defined by the "classical mode" of the largest momentum. 
This way we introduce maximal momentum into the problem. This momentum cut-off 
is the only additional parameter in the method.

Substituting operators for "classical" modes with their corresponding classical amplitudes in 
equation \ref{PlaneHeisenberg} and neglecting all remaining operators we get:
\begin{equation}\label{PlaneGPE}
  \partial_t \alpha_{\textbf{k}}(t)=-i\frac{2\pi^2\hbar}{mL^2}\ k^2\ \alpha_{\textbf{k}}(t)-i\frac{4\pi\hbar aN}{mL^3}
 \sum_{\textbf{k}_1,\ \textbf{k}_2} 
  \alpha^*_{\textbf{k}_1}\alpha_{\textbf{k}_2}\alpha_{\textbf{k}+\textbf{k}_1-\textbf{k}_2}
\end{equation}

We introduce for convenience a unit of energy $\epsilon=4\pi^2\hbar^2/(mL^2)$ and a corresponding unit of time: $\hbar/\epsilon$.
The Eq. \ref{PlaneGPE} can be rewritten in these units in terms of a dimensionless mean-field wave function 
$\psi=\sum_{\textbf{k}} \alpha_{\textbf{k}} e^{-2\pi i\textbf{kr}/L}$ as:
\begin{equation}\label{GPE}
  i \partial_t \psi= -\frac{\Delta}{2} \psi+ g |\psi|^2 \psi,
\end{equation}
where $g=a N/\pi L$ is the interaction strength and the unit for $\psi$ is $L^{-3/2}$.
It is the celebrated Gross-Pitaevskii equation on a grid and can be effectively solved 
by means of split operator method using Fast Fourier Transform (FFT).

\begin{figure}[t]
\includegraphics{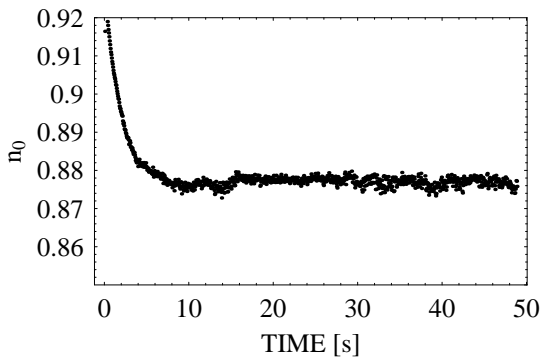}
\caption{Time-evoluiton of a zero momentum component $|\alpha_{0,0,0}(t)|^2$. A thermalization is clearly visible,
as well as rapid fluctuations due to non-linear dynamics. The data have been obtained on a grid 36x36x36 and $g=18.515$.
 } \label{evolution of 000}
\end{figure}

An important feature of Eq.\ref{GPE} is that for almost all initial states $\psi(t=0)$ for a given grid, scattering length $a$ and a 
box size $L$ the evolution leads, after a transient period of time, to a thermodynamically steady 
state that depends only on the energy per particle \cite{Goral1,wiry}. This thermalization is illustrated in Fig. \ref{evolution of 000}, 
where we present the time-evolution of the population of the zero-momentum mode $|\alpha_{0,0,0}(t)|^2$. Thus the number of control parameters for 
equilibrium states is reduced to four: $n_{grid}$, $E$, $L$, $g$.

The wave function $\psi$ stands for both the condensate and the thermal cloud, as opposed to standard 
interpretation where $\psi$ stands for a pure BEC and only the ground state of the GP equation is 
considered. In fact we have not made any distinction between condensed and thermal atoms. 
Instead we use a classic criterion of Onsager and Penrose \cite{Penrose} to identify condensate fraction as a 
dominant eigenvalue of the single particle density matrix. 

A careful reader will notice that $\rho(r,r^\prime)=\psi(r)^*\psi(r^\prime)$ is a pure state and 
has only one eigenvalue. However, typical measurements of BEC involve 
optical techniques, with the exposure time, $\Delta t$, varying from a few up to hundreds of milliseconds. On 
the other hand, the evolution of $\psi$ is very rapid and irregular (note fluctuations in Fig.~\ref{evolution of 000}) 
due to a very short time scales of a nonlinear many-body dynamics. An observation leads to a 
coarse graining and the measured density matrix is of the form:
\begin{equation}\label{time averaging}
  \rho_{aver}(t)=\frac{1}{\Delta t} \int^{t+\Delta t/2}_{t-\Delta t/2} d\tau \ \psi^*(r^\prime,\tau)\psi(r,\tau) \\
\end{equation}

This time averaging procedure destroys the purity of the state by destroying coherence between 
eigenmodes. It also smoothes out the irregularities so that time-averaged density profiles greatly 
resemble experimental photographs \cite{wiry}. This way we can interpret our isolated 
system as a mixed state - a single realization of a quantum system with the properties of a 
measurement process taken into account, which is a case similar to experiments. 

In general, to obtain eigenmodes of the system one needs to diagonalize the time-averaged density matrix 
$\rho_{aver}=\sum_{\textbf{k}} n_{\textbf{k}} \phi_{\textbf{k}}^*(r^{\prime}) \phi_{\textbf{k}}(r)$,
where $\phi_{\textbf{k}}(r)$ are eigenvecors of $\rho_{aver}(r,r^{\prime})$
(see \cite{Goral1} for application of CFA to a harmonic potential). However, the diagonalization is not required for a discussed 3D box 
potential with periodic boundary conditions due to its symmetry \cite{Peter};
$n_{\textbf{k}}$ are just time averaged squares of amplitudes of mode k, $n_{\textbf{k}}=<|\alpha_{\textbf{k}}(t)|^2>$, and
physically they are equal to relative populations of eigenmodes.

This way both dynamic and thermodynamic properties can be simultaneously studied \cite{wiry,Peter,Goral_opt}. 
Still the main problem of the CFA and the goal of this work is to assign results to meaningful physical parameters: the number of particles N and the temperature T. 
To achieve this we now review normal modes of the system - the generalized Bogoliubov quasiparticles, which have been thoroughly studied in \cite{Peter}.
We do this for the sake of completeness of this paper, as we need it to present equipartition of energies in Section \ref{A transition to BEC}, 
which in turn allows us to determine T and N from numerical control parameters.

Typical frequency spectra of time-evolved amplitudes $\alpha_{\textbf{k}}$ are depicted in Fig. ~\ref{typical 
spectra}. For excited modes, they consist of two groups of peaks 
centered at values $\mu \pm \epsilon_{\textbf{k}}$, where $\epsilon_{\textbf{k}}$ is given by the gapless 
Bogoliubov-like formula:

\begin{equation}\label{Bogoliubov energy}
  \epsilon_{\textbf{k}}= \sqrt{(\frac{k^2}{2}+ g n_0)^2-(g n_0)^2}.
\end{equation}

\begin{figure}[t]
\includegraphics[width=0.5\textwidth]{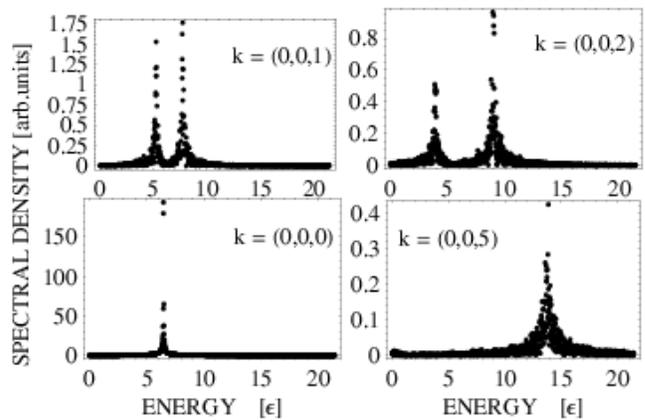}
\caption{
Typical frequency spectra of plane-waves modes. The condensate mode (bottom-left) has a clearly defined energy $\mu$, while spectra of excited modes
are broadened due to interactions.
Note that one group of peaks is suppressed in $(0,0,5)$ mode - the Bogoliubov transformation is practically a unity in this case. 
The parameters of the plot are: $n_{grid}=32$, $g=18.515$, $n_0=86.2\%$. 
} \label{typical spectra}
\end{figure}
and $\mu$ is the energy of the condensate mode. 
For large momentum this formula reduces simply to $k^2/2$. 
With increasing momentum the left group of peaks is suppressed, such that for high momentum modes only one 
group of peaks remains.

\begin{figure}[t]
\begin{center}
\includegraphics[width=0.4\textwidth]{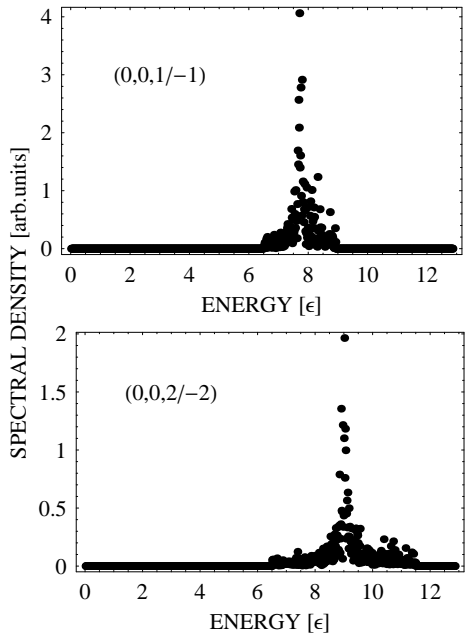}
\caption{
Frequency spectra of quasiparticles - the eigenstates of the system. They have only one central energy, similar to high-momentum plane-wave modes (compare with bottom-right picture in Fig. ~\ref{Bogoliubov energy}). The broadening of the spectrum determines the lifetime of a quasiparticle.
The data was obtained on a grid $n_{grid}=32$ and for the interaction strength $g=18.515$. The condensate population is $86.2\%$.
} \label{quasiparticle spectrum}
\end{center}
\end{figure}

The approximate equations for excited modes couple amplitudes $\alpha_{\textbf{k}}$ with $\alpha^*_{-\textbf{k}}$ \cite{Peter}. 
Provided that spectra of these modes consist of two peaks instead of two groups of peaks, we can perform Bogoliubov transformation
to obtain a quasiparticle amplitude $\delta_{\textbf{k}}$ oscillating with only $one$ frequency $\mu + \epsilon_{\textbf{k}}$.
In our case we deal with two groups of peaks, thus we need to generalize this reasoning. We decompose $\alpha_{\textbf{k}}$ and $\alpha_{-\textbf{k}}$ 
into corresponding pairs of spikes with the frequencies ($\mu + \omega$) and ($\mu-\omega$). By applying the Bogoliubov transformation to each pair
of peaks we arrive at the $(\mu+\omega)$ component of the frequency spectrum of the quasiparticle amplitude: $\delta_{\textbf{k}}(\mu+\omega)$.
Their squared amplitudes are occupations of quasiparticles $n_{\textbf{k}}^{quasi}=|\delta_{\textbf{k}}|^2$.
These quasiparticles represent the eigenmodes of the system with their energies being the central frequency of their spectra (Fig. ~\ref{quasiparticle spectrum}), 
equal to $\mu+\epsilon_{\textbf{k}}$.
The detailed formulas for this procedure are presented in the Appendix.

At the end of this section, let us introduce the cut-off population $N n_{max}$ 
to conveniently confront our method with literature and the assumption of the classical fields approximation. 
We define it as a mean population of "edge" classical modes with momenta greater or equal to 
$\pi\hbar n_{grid}/L$.
Such modes are included in our calculations because we work with rectangular grid.   
The population $N n_{max}$ is the population cut-off that separates classical (mean-field) and quantum (neglected) modes 
and quantitatively specifies the term "sufficiently large population".
It can be used as an alternative numerical control parameter instead of the grid size $n_{grid}$ \cite{Peter,Goral_opt}, 
as $N n_{max}$ is defined by the grid for given $L$, $g$ and $E$.

\section{A transition to BEC. A distribution of energies.}\label{A transition to BEC}

In this section we analyze properties of the system in thermal equilibrium above and below the critical energy.
We use the equipartition of energies in eigenmodes of the system 
to determine the dimensionless temperature per particle $\tau$. Finally, we discuss in detail the inconsistency emerging 
from the free choice of the grid for numerical calculations.

\begin{figure}[t]
\includegraphics{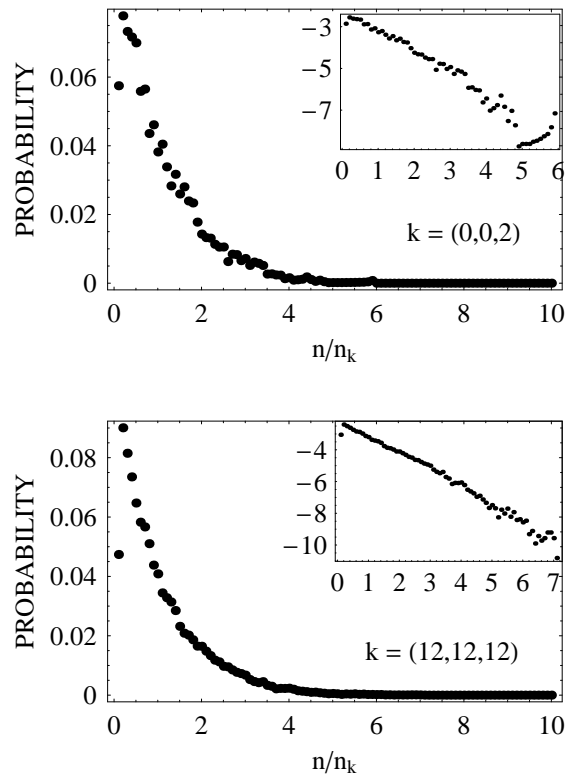}
\caption{
A probability distribution of occupations of excited modes. 
Plots in insets are in logarithmic scale to outline the observed exponential trend.
The parameters are: $n_{grid}=32$, $n_0=86.2\%$ and $g=18.515$.
}
\label{histogramy term}
\end{figure}

First, we inspect the behavior of excited modes in the state of equilibrium. 
We define $p_{\textbf{k}}(n)$ to be a probability of finding 
the population of quasiparticle $n_{\textbf{k}}^{quasi}$ equal to $n$. 
It is calculated by counting all events where $|\delta_{\textbf{k}}|^2$ is close to $n$ during a simulation. 
Results for $p_{\textbf{k}}$ are presented in Fig.~\ref{histogramy term}. 
They reveal roughly exponential distribution of populations of thermal modes 
and provide a strong argument that we observe the thermalization in the system.
Thus one can treat a single excited
mode as being in thermal equilibrium with a reservoir consisting of the rest of the system.

\begin{figure}[t]
\includegraphics[width=0.35\textwidth]{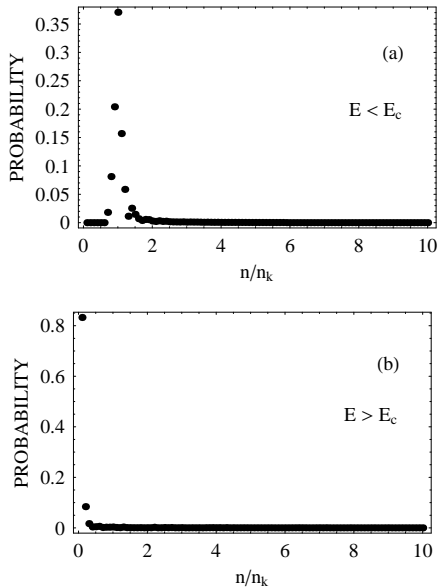}
\caption{
A probability distribution of occupations of the condensate mode: (a) below $E_c$ ($g=18.515$, $n_g$=48, $n_0=7.2\%$),
(b) above $E_c$ ($n_g=50$, $g=18.515$, $E=300$).
}
\label{histogramy000}
\end{figure}

For comparison, similar calculations of $p_{0,0,0}(n)$ performed for the condensate mode $\alpha_{0,0,0}$ 
reveal a phase transition into BEC (\cite{wiry, Scully, Wilkens}).
Below the critical energy the distribution is peaked around a non-zero condensate population
(Fig.~\ref{histogramy000}a), while above $E_c$ it remains exponential (Fig.~\ref{histogramy000}b).

\begin{figure}[t]
\includegraphics[width=0.4\textwidth]{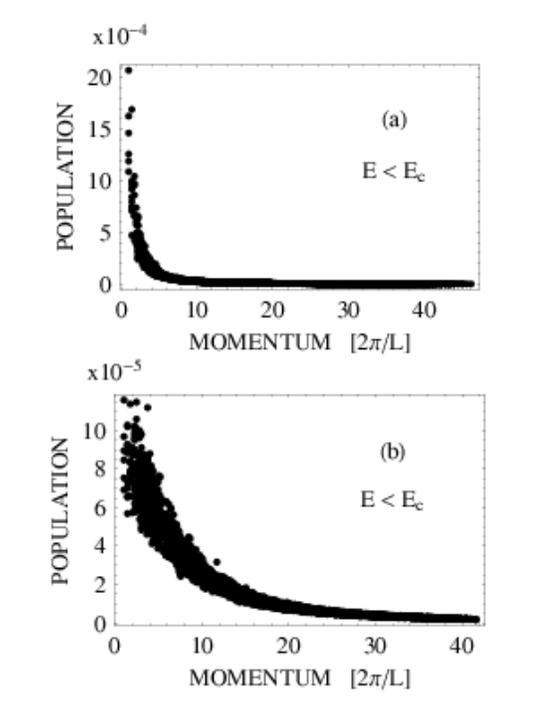}
\caption{
Populations of thermal modes versus their momentum: below (a) and above the critical energy (b).
Parameters are: (a) $n_{grid}=54$, $E=215$, $n_0=0.3\%$, $g=18.515$, 
(b) $n_{grid}=50$, $E=225$, $g=18.515$.
Note how a very small fraction of condensed atoms ($0.3\%$) can influence the whole system.
}\label{distrib}
\end{figure}

Let us now consider a distribution of populations of thermal modes $n_{k}$.
We have found that this feature differs for condensed and uncondensed systems (see Fig.~\ref{distrib})
and defines the critical energy more accurately than the change in $p_{0,0,0}(n)$.
Below the $E_c$ the distribution is proportional to $k^{-2}$ and this agrees with the low temperature
limit for the population distribution of an ideal Bose gas.
Above the critical energy the distribution abruptly changes to $\exp(-k)$. 

The distribution of populations is closely related to the equipartition of energies occurring in eigenmodes of the system:
\begin{equation}\label{equipartition:relation}
N n_k \epsilon_k=k_BT.
\end{equation}
This important concept, introduced to the CFA in \cite{DavisBurnett} and expanded in \cite{Peter}, 
allows one to easily determine the dimensionless temperature per particle $\tau$ - 
the essential parameter for our method.

\begin{figure}[t]
\includegraphics[width=0.4\textwidth]{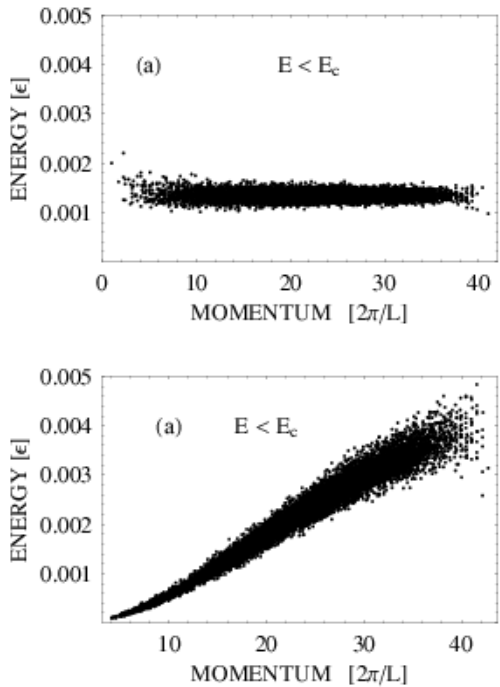}
\caption{
The energy confined in excited modes ($\epsilon_{\textbf{k}} n_{\textbf{k}}$) versus the momentum for $g=18.515$:
(a) below $E_c$, $n_{grid}=48$, $E=163.6$ and (b) above $E_c$, $n_{grid}=50$, $E=300$. 
Only below the critical temperature ($n_0=7.2\%$) one observes the equipartition of energy. 
Note that the condensate mode does not satisfy the equipartition.
}\label{equipartition}
\end{figure}

We find that the equipartition occurs only below the critical energy, 
thus limiting its application to calculate the temperature to below the critical temperature $T_c$. 
One can notice this easily for high-momentum modes (see Fig.~\ref{equipartition}a), 
where $\epsilon_k=\hbar^2 k^2/2m$ and the equipartition reduces to $n_k k^2=const$.
Clearly, the exponential population distribution occurring above the critical energy $E_c$ cannot satisfy this relation
(Fig.~\ref{equipartition}b).

\begin{figure}[t]
\includegraphics{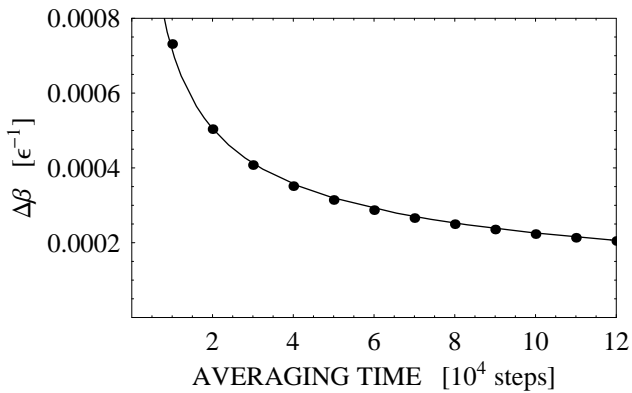}
\caption{
Fluctuations of an inverse dimensionless temperature per particle $\beta=\tau^{-1}$ as a function of the number of averaging time-steps.
The data is obtained on a grid $n_{grid}=48$ and the condensate population is $n_0=7.2\%$.
}\label{clt}
\end{figure}

Before we make use of the equipartition to obtain $\tau$, let us present yet another argument supporting
the quality of observed equilibrium - the fluctuations of energy, visible in Fig.~\ref{equipartition}a. 
A system which is truly thermalized experiences such fluctuations, 
but they vanish with the averaging time as $\Delta t^{-1/2}$,
according to the Central Limit Theorem. 
Indeed, the agreement of results with the theorem is excellent, as can be seen in Fig.~\ref{clt}. 

\begin{figure}[t]
\includegraphics[width=0.4\textwidth]{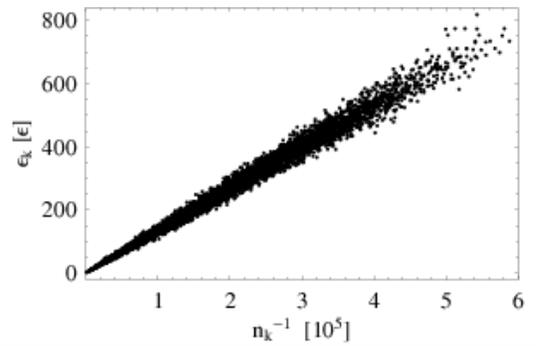}
\caption{
The energies of quasiparticles versus their inverse populations depicted for a condensed system ($n_0=7.2\%$, $n_{grid}=48$, $E=163.6$, $g=18.515$). A slope of the linear fit to the data is the dimensionless temperature per particle $\tau$.
}\label{temperature}
\end{figure}

Now we can determine the dimensionless temperature per particle $\tau=k_B T/\epsilon N$. 
The equipartition can be written in a form independent of $N$ and $L$ as:
\begin{equation}\label{equipartition:formula}
 \epsilon_{\textbf{k}}= \tau\frac{1}{n_{\textbf{k}}}
\end{equation}
For different condensate fractions $n_0$ this relation is depicted in Figure ~\ref{temperature}. 
The parameter $\tau$ is a slope of the best linear fit to numerical data \cite{Peter}.
It depends on three numerical parameters: the energy per particle $E$, the interaction $g$ and the grid size $n_{grid}$, 
such that if $g$ is set, then a pair ($n_0$, $\tau$) is unequivocally determined by $E$ and $n_{grid}$.

There are two common ways of specifying the classical fields approximation in the literature. 
One of them \cite{Peter,wiry} defines quantitatively the assumption of
"sufficiently" large population of modes treated classically by arbitrarily setting the value
of the population cut-off $Nn_{max}$.
The other method \cite{Davis,DavisBurnett} assumes that numerical results represent only a fraction of a bigger system;
the population cut-off is large (i.e. greater than 15) and external atoms are approximated by an ideal gas.

\begin{figure}[t]
\includegraphics[width=0.5\textwidth]{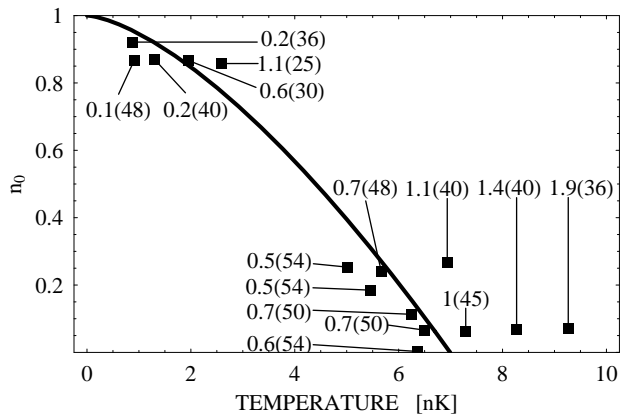}
\caption{
The condensate fraction versus the temperature for various numerical grids. The interaction strength $g$ is $18.515$ and the number of atoms $N$ is set to $235000$. Numbers describing the points in the figure are the population cut-off $N n_{max}$ (left numbers) and the grid size $n_{grid}$ (numbers in brackets). The population cut-off is known only approximately due to large fluctuations in highly excited modes. The optimal value of the cut-off, for which the numerical results agree with the corresponding ideal Bose gas is ca. $0.6-0.7$ in this case. 
}
\label{temperature vs cut-off}
\end{figure}

The major concern of the CFA is the dependence of results on the 
particular choice of the grid, as it does not have any physical meaning.
It is illustrated in Fig. ~\ref{temperature vs cut-off}, 
which depicts the condensate fraction $n_0$ versus the temperature $T=\tau\epsilon N/L$
obtained for various grids $n_{grid}$ and energies per particle $E$, while $g$ and $L$ are set. 
We have assumed $N$ to be constant (235000) so that we can determine $T$ and the population cut-off $N n_{max}$.
A solid curve represents a relation $n_0(T)$ of the ideal Bose gas with the same N and L. 
For interacting Bose gas this relation would be only slightly modified, as
for small scattering length corrections to the critical temperature are known to be small \cite{Alber,poprawkidoTc,poprawkidoTc2}. 

One can clearly see that by changing the numerical grid it is possible to obtain a relatively broad range of
temperatures $T$ (and $\tau$) for a single condensate fraction $n_0$.
The grid that reproduces the result for the ideal Bose gas at one value of $n_0$,
fails to do so for different condensate fractions.
On the other hand there exists an optimal value of the population cut-off $Nn_{max}$ (ca. $0.6-0.7$)
for which the results match the ideal Bose gas both for low temperatures or close to $T_c$. 
If $Nn_{max}$ differs from the optimal value not too much, 
the results follow a curve which is shifted with respect to the ideal gas 
(see results in \cite{Peter}, where the population cut-off is arbitrarily set to 1). 
Such results are still reliable, although their accuracy is reduced by several tens of per cents.
However, the optimal value of the population cut-off $Nn_{max}$ changes with the interaction strength $g$, so that there
is no universal value that satisfies all control parameters.
And as its physical interpretation can be qualitative at most,
the population cut-off is still a free parameter of the method, supplementary to the grid size $n_{grid}$.

To summarize, we have analyzed the properties of the thermal equilibrium of the system and 
shown difficulties arising from the existence of artificial parameter ($Nn_{max}$ or $n_{grid}$) in the classical fields approximation.
We have based the quality of results on the comparison of the temperature calculated from equipartition 
with the corresponding temperature of the ideal Bose gas. 
In the next section we build on this concept to obtain a method which is free of the presented vulnerabilities
and yet describes the whole gas within the classical fields in a simple and consistent way.

\section{Eliminating the cut-off}\label{Disposing of cut-off}

We base this approach on an assumption 
that numerical results ($n_0$, $\tau$) represent a certain entire Bose gas with the condensate fraction $n_0$.
We impose a condition that the temperature obtained from the equipartition $T=\frac{4\pi^2\hbar^2N}{mL^2}\ \tau$
matches the temperature of the corresponding Bose gas, approximated with a formula for an 
ideal Bose gas:
\begin{equation}\label{T:ideal}
 T^{ideal}(n_0,N,L)=\frac{2\pi\hbar^2}{mk_B}\zeta^{-2/3}(3/2) (1-n_0)^{2/3} \frac{N^{2/3}}{L^2}, 
\end{equation}
where $\zeta$ is the Riemann function.

To justify this approximation, note that for low scattering lengths the critical temperature of an interacting Bose gas 
is very close to that of 
of an ideal Bose gas, although there is a significant disagreement between various
authors over its precise value (see for instance \cite{Alber} for calculations of the corrections). 
For typical systems with the atomic density of $1.8 \  10^{13} cm^{-3}$ 
and the scattering length $a=5.8 nm$ the correction
to $T_c$ is about $4.5\%$. 

Because formulas for both temperatures scale differently with the number of particles, 
we can always satisfy our assumption by taking:
\begin{equation}\label{N}
  N= \frac{(1-n_0)^2}{\zeta^{2}(\frac{3}{2})\ \pi^{3}\tau^{3}}.
\end{equation}

This way we determine the number of particles $N$ and the temperature $T$ of the system from  
the condition that has a clearly defined physical meaning.
The remaining physical control parameter - the scattering length $a$ 
can be now easily calculated from the value of the interaction strength $g=a N/\pi L$ used in numerical calculations.

The main advantage of this approach is that we do not set arbitrarily any artificial parameter, because
$N$ is implicitly dependent on the grid size $n_{grid}$.
Thus problems arising from a freedom of choice of the numerical grid, such as outlined in Section \ref{A transition to BEC}, 
do not occur anymore.

We perform a rough check of consistency of our result by comparing the kinetic energy per particle 
obtained from the simulation with the kinetic energy per particle of an ideal Bose gas, which is given by the 
formula:
\begin{equation}\label{Ekin:ideal}
\frac{E_{kin}}{N}=\frac{3}{2}\pi^{\frac{3}{2}} \epsilon\:
\zeta(\frac{5}{2}) 
N^{\frac{3}{2}} \tau^{\frac{5}{2}}.
\end{equation}

The results show an agreement within $40\%$
, what is acceptable due to the very approximate nature of this comparison.

Next, we calculate the cut-off population $Nn_{max}$. It represents the optimal value of the cut-off for chosen 
condensate fraction $n_0$ and interaction strength $g$. 
It is remarkable that, even though we change the grid and adjust the energy per particle such that 
$n_0$ remains constant, our procedure alters the number of particles in such a way, that the cut-off 
remains relatively fixed (approx. $0.7$ for $g=18.515$). In general, it depends only on the interaction strength, and this agrees with our 
previous observations (see Chapter \ref{A transition to BEC}).

The relative invariance of the population cut-off with respect to the grid size $n_{grid}$ and the
condensate fraction $n_0$ can be understood by considering a simplified model of a system with equipartition
of energies in plane waves instead of proper eigenmodes. 
Such system (considered earlier in \cite{Davis}) corresponds to the case of $g=0$, thus it 
cannot be obtained within the CFA, which relies on interactions for thermalization.
We calculate the number of particles again by comparing the dimensionless temperature,
given by the relation $\tau=(1-n_0)/(\sum_{\textbf{k}}{\frac{1}{k^2}})$ with the temperature of an ideal Bose gas.
The cut-off population for this system is given by the formula:
\begin{equation}\label{cut-off:grid}
  N n_{max}=\frac{(\sum_{\textbf{k}}{\frac{1}{k^2}})^2}{\pi^3 \zeta^2(\frac{3}{2})}\ \frac{\sum_{|\textbf{k}|\ge \pi n_{grid}/L}{\frac{1}{k^2}}}
{\sum_{|\textbf{k}|\ge \pi n_{grid}/L}1}.
\end{equation}
Similar to the interacting case, this population cut-off does not depend on the condensate fraction $n_0$.
For the grid size $n_{grid}=16$ it equals $0.69$, while at $n_{grid}=128$ it is $0.80$. 
This change in the population cut-off can be attributed to the box shape of the grid. Indeed, if we
assume a spherical grid and approximate summations in Eq. \ref{cut-off:grid} with integrals, then the result 
($Nn_{max}=16\pi^{-1}\zeta^{-2}(3/2)=0.75$) is independent also from the grid size.

Note that the presented approach does not depend on the value of the population cut-off (or any other artificial parameter).
This parameter serves only for comparison with the assumption of "sufficiently large" populations of classical modes.
It can be surprising that the optimal cut-off population comes out so small,
but on the other hand it shows that almost whole gas is confined in the classical modes. 
An estimate based upon approximating external modes with an ideal gas and gluing both gases with a single
temperature and the same cut-off population on the border yields about $97-99\%$ atoms inside the CFA grid.
This result is consistent with our assumption that we describe the entire system.

At the end, we also discuss two drawbacks of the presented approach.
First, it does not allow to study the shift of critical temperature upon the scattering length, as we
fix the temperature to the value of an ideal gas. 
The second limitation is that one can apply this method only below the $T_c$, as we use equipartition of
energy which occurs in a condensed phase only.
However, we are very uncertain if the CFA works above $T_c$ (at least in 3D box), as the observed exponential 
population distribution of excited modes does not agree with the distribution of an ideal Bose gas.
Also, the CFA has not been constructed for this purpose.

\section{Conclusions}\label{Conclusions}

We have presented the version of the Classical Fields Approximation without free parameters, 
which finally allows one to easily and consistently present results versus real-world parameters, 
like the temperature and the number of particles. Most important, the accuracy of these results
is unbiased by the choice of the numerical grid - a major concern of former applications of the CFA method, as we rely on the 
condition with a clear physical meaning rather than arbitrarily set some artificial parameter.

We have calculated the optimal population cut-off limiting the classical and quantum regimes in the CFA 
and found that it almost does not depend on the condensate fraction or the grid size.
We explain this behavior
with a simplified model of a non-interacting gas on a grid and find considerable agreement.
As a conclusion we validate the previous use of the fixed cut-off classical fields approximation \cite{Peter}, 
although we point to the hardly controlled accuracy of such results.

Moreover we have shown that one can describe the entire gas within the classical fields, as opposed to an approach presented
in \cite{Davis}, where the CFA is considered to describe only a small part of the whole system. However, from the obtained value
of the population cut-off we conclude that the classical 
fields approximation is exploited to its limits of applicability.

We have also presented the detailed analysis of the thermal equilibrium of the system. We have reported on the rapid change 
in the distribution of populations of the excited modes which marks the phase transition into BEC in the classical fields approximation.
We have shown that the equipartition of energies in eigenmodes of the system occurs in a condensed phase only,
which limits the use of our method to temperatures below the critical one.

\begin{acknowledgments}

The authors acknowledge support of the Polish Ministry of Scientific Research and Information Technology under Grant No. PBZ-MIN-008/P03/2003. 
The results have been obtained using computers at the Interdisciplinary Centre for
Mathematical and Computational Modeling of Warsaw University.

\end{acknowledgments}

\appendix*
\section{A Bogolubov transformation}

We present here the derivation of formulas for the generalized Bogoliubov quasiparticles, which have been described in Section
\ref{The Method}. 
The observed spectra of amplitudes of excited modes consist of two groups of peaks (see Fig. ~\ref{typical 
spectra}), thus we describe the amplitude of mode $\textbf{k}$ as:
\begin{equation}\label{peak decomposition}
  \alpha_{\textbf{k}}(t)=\sum_{\omega}\beta_{\textbf{k}}(\omega)e^{-i(\mu-\omega)t} + \gamma_{\textbf{k}}(\omega)e^{-i(\mu+\omega)t},
\end{equation}
where the coefficients $\beta_{\textbf{k}}(\omega)$ and $\gamma_{\textbf{k}}(\omega)$ are obtained from simulation.

As the approximate equations for excited modes couple amplitudes $\alpha_{\textbf{k}}$ with $\alpha^*_{-\textbf{k}}$ \cite{Peter},
we can identify the corresponding pairs of spikes $\alpha_{\textbf{k}}(\mu+\omega)$ and $\alpha_{-\textbf{k}}(\mu-\omega)$: they are
$\gamma_{\textbf{k}}(\omega)$ and $\beta_{-\textbf{k}}(\omega)$, respectively.
We apply a Bogoliubov transform to each such pair to obtain a single spike having only one frequency.
In order to do so, we introduce a new quasiparticle amplitude $\delta_{\textbf{k}}$, defined as follows:
\begin{equation}\label{quasiparticle:peak}
\delta_{\textbf{k}}(\mu+\omega)=
U_{\textbf{k}}(\omega)\alpha_{\pm \textbf{k}}(\mu+\omega)+e^{-2i\mu t}V_{\textbf{k}}(\omega)\alpha^*_{-\textbf{k}}(\mu-\omega),
\end{equation}
where $U_{\textbf{k}}(\omega)$ and $V_{\textbf{k}}(\omega)$ satisfy the condition:
\begin{equation}\label{quasiparticle:condition}
 |U_{\pm \textbf{k}}(\omega)|^2-|V_{\pm \textbf{k}}(\omega)|^2=1 
\end{equation}

Substituting \ref{peak decomposition} into \ref{quasiparticle:peak} we get:
\begin{equation}\label{quasiparticle:peak2}
\begin{split}
\delta_{\textbf{k}}(\mu+\omega)=
(U_{\textbf{k}}(\omega)\ \beta_{\textbf{k}}(\omega)+V_{\textbf{k}}(\omega)\ \gamma^*_{-\textbf{k}}(\omega)) e^{-i(\mu-\omega)t} +\\
(U_{\textbf{k}}(\omega)\ \gamma_{\textbf{k}}(\omega)+V_{\textbf{k}}(\omega)\ \beta^*_{-\textbf{k}}(\omega)) e^{-i(\mu+\omega)t}.
\end{split}
\end{equation}

We want $\delta_{\textbf{k}}(\mu+\omega)$ to be a component of a quasiparticle with only single positive 
energy, thus we impose a condition:
\begin{equation}\label{quasiparticle:condition2}
 U_{\textbf{k}}(\omega)\ \beta_{\textbf{k}}(\omega) +V_{\textbf{k}}(\omega)\ \gamma^*_{-\textbf{k}}(\omega)=0. 
\end{equation}

From eq. \ref{quasiparticle:condition} and \ref{quasiparticle:condition2} we obtain:
\begin{align}\label{quasiparticle:uv}
  U_{\textbf{k}}(\omega)=\frac{|\gamma_{-\textbf{k}}(\omega)|}{\sqrt{|\gamma_{-\textbf{k}}(\omega)|^2-|\beta_{\textbf{k}}(\omega)|^2}}\ e^{-i Arg(\beta_{\textbf{k}}(\omega))} \\
  V_{\textbf{k}}(\omega)=-\frac{|\beta_{\textbf{k}}(\omega)|}{\sqrt{|\gamma_{-\textbf{k}}(\omega)|^2-|\beta_{\textbf{k}}(\omega)|^2}}\ e^{i Arg(\gamma_{-\textbf{k}}(\omega))}.
\end{align}
	
Finally we arrive at the $\mu+\omega$ component of the quasiparticle amplitude composed of modes $\textbf{k}$ and $-\textbf{k}$:
\begin{equation}\label{quasiparticle:amplitude}
\begin{split}
\delta_{\textbf{k}}(\mu+\omega)= \frac{e^{-i (\mu+\omega) t}}{\sqrt{|\gamma_{-\textbf{k}}(\omega)|^2-|\beta_{\textbf{k}}(\omega)|^2}}\\
( |\gamma_{\textbf{k}}(\omega)\gamma_{-\textbf{k}}(\omega)| e^{i(Arg(\gamma_{\textbf{k}}(\omega))-Arg(\beta_{\textbf{k}}(\omega)))}-\\
|\beta_{\textbf{k}}(\omega)\beta_{-\textbf{k}}(\omega)| e^{i(Arg(\gamma_{-\textbf{k}}(\omega))-Arg(\beta_{-\textbf{k}}(\omega)))}\ )
\end{split}
\end{equation}
Spectra of these excitations are centered on a single value (Fig. ~\ref{quasiparticle 
spectrum}), similar to high momentum plane-wave modes. They can be regarded as normal modes 
of the system oscillating with approximately single frequency, 
while the width of the spectrum is related to the life-time of a quasiparticle \cite{Peter}. 
\\


\begin{thebibliography}{21}
\expandafter\ifx\csname natexlab\endcsname\relax\def\natexlab#1{#1}\fi
\expandafter\ifx\csname bibnamefont\endcsname\relax
  \def\bibnamefont#1{#1}\fi
\expandafter\ifx\csname bibfnamefont\endcsname\relax
  \def\bibfnamefont#1{#1}\fi
\expandafter\ifx\csname citenamefont\endcsname\relax
  \def\citenamefont#1{#1}\fi
\expandafter\ifx\csname url\endcsname\relax
  \def\url#1{\texttt{#1}}\fi
\expandafter\ifx\csname urlprefix\endcsname\relax\def\urlprefix{URL }\fi
\providecommand{\bibinfo}[2]{#2}
\providecommand{\eprint}[2][]{\url{#2}}

\bibitem[{\citenamefont{Anderson et~al.}(1995)\citenamefont{Anderson, Ensher,
  Wieman, and Cornell}}]{BEC1}
\bibinfo{author}{\bibfnamefont{M.~H.} \bibnamefont{Anderson}},
  \bibinfo{author}{\bibfnamefont{J.~R.} \bibnamefont{Ensher}},
  \bibinfo{author}{\bibfnamefont{C.}~\bibnamefont{Wieman}}, \bibnamefont{and}
  \bibinfo{author}{\bibfnamefont{E.}~\bibnamefont{Cornell}},
  \bibinfo{journal}{Science} \textbf{\bibinfo{volume}{269}},
  \bibinfo{pages}{198} (\bibinfo{year}{1995}).

\bibitem[{\citenamefont{Davis et~al.}(1995)\citenamefont{Davis, Mewes, Andrews,
  van Druten, Durfee, Kurn, and Ketterle}}]{BEC2}
\bibinfo{author}{\bibfnamefont{K.~B.} \bibnamefont{Davis}},
  \bibinfo{author}{\bibfnamefont{M.-O.} \bibnamefont{Mewes}},
  \bibinfo{author}{\bibfnamefont{M.~R.} \bibnamefont{Andrews}},
  \bibinfo{author}{\bibfnamefont{N.~J.} \bibnamefont{van Druten}},
  \bibinfo{author}{\bibfnamefont{D.~S.} \bibnamefont{Durfee}},
  \bibinfo{author}{\bibfnamefont{D.~M.} \bibnamefont{Kurn}}, \bibnamefont{and}
  \bibinfo{author}{\bibfnamefont{W.}~\bibnamefont{Ketterle}},
  \bibinfo{journal}{Phys. Rev. Lett.} \textbf{\bibinfo{volume}{75}},
  \bibinfo{pages}{3969} (\bibinfo{year}{1995}).

\bibitem[{\citenamefont{Chu}(1998)}]{BEC3}
\bibinfo{author}{\bibfnamefont{S.}~\bibnamefont{Chu}}, \bibinfo{journal}{Rev.
  Mod. Phys.} \textbf{\bibinfo{volume}{70}}, \bibinfo{pages}{685}
  (\bibinfo{year}{1998}).

\bibitem[{\citenamefont{Jackson and Zaremba}(2002)}]{two-gas}
\bibinfo{author}{\bibfnamefont{B.}~\bibnamefont{Jackson}} \bibnamefont{and}
  \bibinfo{author}{\bibfnamefont{E.}~\bibnamefont{Zaremba}},
  \bibinfo{journal}{Phys. Rev. Lett.} \textbf{\bibinfo{volume}{88}},
  \bibinfo{pages}{180402} (\bibinfo{year}{2002}).

\bibitem[{\citenamefont{Hutchinson et~al.}(1997)\citenamefont{Hutchinson,
  Zaremba, and Griffin}}]{two-gas2}
\bibinfo{author}{\bibfnamefont{D.~A.~W.} \bibnamefont{Hutchinson}},
  \bibinfo{author}{\bibfnamefont{E.}~\bibnamefont{Zaremba}}, \bibnamefont{and}
  \bibinfo{author}{\bibfnamefont{A.}~\bibnamefont{Griffin}},
  \bibinfo{journal}{Phys. Rev. Lett.} \textbf{\bibinfo{volume}{78}},
  \bibinfo{pages}{1842} (\bibinfo{year}{1997}).

\bibitem[{\citenamefont{Khawaja and Stoof}(2000)}]{two-gaszmag2}
\bibinfo{author}{\bibfnamefont{U.~A.} \bibnamefont{Khawaja}} \bibnamefont{and}
  \bibinfo{author}{\bibfnamefont{H.~T.~C.} \bibnamefont{Stoof}},
  \bibinfo{journal}{Phys. Rev. A} \textbf{\bibinfo{volume}{62}},
  \bibinfo{pages}{053602} (\bibinfo{year}{2000}).

\bibitem[{\citenamefont{Dodd et~al.}(1998)\citenamefont{Dodd, Edwards, Clark,
  and Burnett}}]{two-gas3popov}
\bibinfo{author}{\bibfnamefont{R.~J.} \bibnamefont{Dodd}},
  \bibinfo{author}{\bibfnamefont{M.}~\bibnamefont{Edwards}},
  \bibinfo{author}{\bibfnamefont{C.~W.} \bibnamefont{Clark}}, \bibnamefont{and}
  \bibinfo{author}{\bibfnamefont{K.}~\bibnamefont{Burnett}},
  \bibinfo{journal}{Phys. Rev. A} \textbf{\bibinfo{volume}{57}},
  \bibinfo{pages}{R32} (\bibinfo{year}{1998}).

\bibitem[{\citenamefont{Sinatra et~al.}(2002)\citenamefont{Sinatra, Lobo, and
  Castin}}]{Castin}
\bibinfo{author}{\bibfnamefont{A.}~\bibnamefont{Sinatra}},
  \bibinfo{author}{\bibfnamefont{C.}~\bibnamefont{Lobo}}, \bibnamefont{and}
  \bibinfo{author}{\bibfnamefont{Y.}~\bibnamefont{Castin}},
  \bibinfo{journal}{J. Phys. B} \textbf{\bibinfo{volume}{35}},
  \bibinfo{pages}{3599} (\bibinfo{year}{2002}).

\bibitem[{\citenamefont{Davis and Morgan}(2003)}]{Davis}
\bibinfo{author}{\bibfnamefont{M.~J.} \bibnamefont{Davis}} \bibnamefont{and}
  \bibinfo{author}{\bibfnamefont{S.~A.} \bibnamefont{Morgan}},
  \bibinfo{journal}{Phys. Rev. A} \textbf{\bibinfo{volume}{68}},
  \bibinfo{pages}{053615} (\bibinfo{year}{2003}).

\bibitem[{\citenamefont{Kagan and Svistunov}(1997)}]{CFAinni}
\bibinfo{author}{\bibfnamefont{Y.}~\bibnamefont{Kagan}} \bibnamefont{and}
  \bibinfo{author}{\bibfnamefont{B.~V.} \bibnamefont{Svistunov}},
  \bibinfo{journal}{Phys. Rev. Lett.} \textbf{\bibinfo{volume}{79}},
  \bibinfo{pages}{3331} (\bibinfo{year}{1997}).

\bibitem[{\citenamefont{G{\'o}ral et~al.}(2002)\citenamefont{G{\'o}ral, Gajda,
  and Rz{\c a}{\.z}ewski}}]{Goral1}
\bibinfo{author}{\bibfnamefont{K.}~\bibnamefont{G{\'o}ral}},
  \bibinfo{author}{\bibfnamefont{M.}~\bibnamefont{Gajda}}, \bibnamefont{and}
  \bibinfo{author}{\bibfnamefont{K.}~\bibnamefont{Rz{\c a}{\.z}ewski}},
  \bibinfo{journal}{Phys. Rev. A} \textbf{\bibinfo{volume}{66}},
  \bibinfo{pages}{051602} (\bibinfo{year}{2002}).

\bibitem[{\citenamefont{Brewczyk et~al.}(2004)\citenamefont{Brewczyk, Borowski,
  Gajda, and Rz{\c a}{\.z}ewski}}]{Peter}
\bibinfo{author}{\bibfnamefont{M.}~\bibnamefont{Brewczyk}},
  \bibinfo{author}{\bibfnamefont{P.}~\bibnamefont{Borowski}},
  \bibinfo{author}{\bibfnamefont{M.}~\bibnamefont{Gajda}}, \bibnamefont{and}
  \bibinfo{author}{\bibfnamefont{K.}~\bibnamefont{Rz{\c a}{\.z}ewski}},
  \bibinfo{journal}{J. Phys. B.} \textbf{\bibinfo{volume}{37}},
  \bibinfo{pages}{2725} (\bibinfo{year}{2004}).

\bibitem[{\citenamefont{G{\'o}ral et~al.}(2001)\citenamefont{G{\'o}ral, Gajda,
  and Rz{\c a}{\.z}ewski}}]{Goral_opt}
\bibinfo{author}{\bibfnamefont{K.}~\bibnamefont{G{\'o}ral}},
  \bibinfo{author}{\bibfnamefont{M.}~\bibnamefont{Gajda}}, \bibnamefont{and}
  \bibinfo{author}{\bibfnamefont{K.}~\bibnamefont{Rz{\c a}{\.z}ewski}},
  \bibinfo{journal}{Opt. Express} \textbf{\bibinfo{volume}{8}},
  \bibinfo{pages}{92} (\bibinfo{year}{2001}).

\bibitem[{\citenamefont{Schmidt et~al.}(2003)\citenamefont{Schmidt, G{\'o}ral,
  Floegel, Gajda, and Rz{\c a}{\.z}ewski}}]{wiry}
\bibinfo{author}{\bibfnamefont{H.}~\bibnamefont{Schmidt}},
  \bibinfo{author}{\bibfnamefont{K.}~\bibnamefont{G{\'o}ral}},
  \bibinfo{author}{\bibfnamefont{F.}~\bibnamefont{Floegel}},
  \bibinfo{author}{\bibfnamefont{M.}~\bibnamefont{Gajda}}, \bibnamefont{and}
  \bibinfo{author}{\bibfnamefont{K.}~\bibnamefont{Rz{\c a}{\.z}ewski}},
  \bibinfo{journal}{J. Opt. B} \textbf{\bibinfo{volume}{5}},
  \bibinfo{pages}{S96} (\bibinfo{year}{2003}).

\bibitem[{\citenamefont{Davis et~al.}(2002)\citenamefont{Davis, Morgan, and
  Burnett}}]{DavisBurnett}
\bibinfo{author}{\bibfnamefont{M.~J.} \bibnamefont{Davis}},
  \bibinfo{author}{\bibfnamefont{S.~A.} \bibnamefont{Morgan}},
  \bibnamefont{and} \bibinfo{author}{\bibfnamefont{K.}~\bibnamefont{Burnett}},
  \bibinfo{journal}{Phys. Rev. A} \textbf{\bibinfo{volume}{66}},
  \bibinfo{pages}{053618} (\bibinfo{year}{2002}).

\bibitem[{\citenamefont{Penrose and Onsager}(1956)}]{Penrose}
\bibinfo{author}{\bibfnamefont{O.}~\bibnamefont{Penrose}} \bibnamefont{and}
  \bibinfo{author}{\bibfnamefont{L.}~\bibnamefont{Onsager}},
  \bibinfo{journal}{Phys. Rev.} \textbf{\bibinfo{volume}{104}},
  \bibinfo{pages}{576} (\bibinfo{year}{1956}).

\bibitem[{\citenamefont{Kocharovsky et~al.}(2000)\citenamefont{Kocharovsky,
  Kocharovsky, and Scully}}]{Scully}
\bibinfo{author}{\bibfnamefont{V.~V.} \bibnamefont{Kocharovsky}},
  \bibinfo{author}{\bibfnamefont{V.~V.} \bibnamefont{Kocharovsky}},
  \bibnamefont{and} \bibinfo{author}{\bibfnamefont{M.~O.}
  \bibnamefont{Scully}}, \bibinfo{journal}{Phys. Rev. A}
  \textbf{\bibinfo{volume}{61}}, \bibinfo{pages}{053606}
  (\bibinfo{year}{2000}).

\bibitem[{\citenamefont{Illuminati et~al.}(1999)\citenamefont{Illuminati,
  Navez, and Wilkens}}]{Wilkens}
\bibinfo{author}{\bibfnamefont{F.}~\bibnamefont{Illuminati}},
  \bibinfo{author}{\bibfnamefont{P.}~\bibnamefont{Navez}}, \bibnamefont{and}
  \bibinfo{author}{\bibfnamefont{M.}~\bibnamefont{Wilkens}},
  \bibinfo{journal}{J. Phys. B} \textbf{\bibinfo{volume}{31}},
  \bibinfo{pages}{L461} (\bibinfo{year}{1999}).

\bibitem[{\citenamefont{Alber}(2001)}]{Alber}
\bibinfo{author}{\bibfnamefont{G.}~\bibnamefont{Alber}},
  \bibinfo{journal}{Phys. Rev. A} \textbf{\bibinfo{volume}{63}},
  \bibinfo{pages}{023613} (\bibinfo{year}{2001}).

\bibitem[{\citenamefont{Arnold and Moore}(2001)}]{poprawkidoTc}
\bibinfo{author}{\bibfnamefont{P.}~\bibnamefont{Arnold}} \bibnamefont{and}
  \bibinfo{author}{\bibfnamefont{G.}~\bibnamefont{Moore}},
  \bibinfo{journal}{Phys. Rev. Lett.} \textbf{\bibinfo{volume}{87}},
  \bibinfo{pages}{120401} (\bibinfo{year}{2001}).

\bibitem[{\citenamefont{Kashurnikov et~al.}(2001)\citenamefont{Kashurnikov,
  Prokof'ev, and Svistunov}}]{poprawkidoTc2}
\bibinfo{author}{\bibfnamefont{V.~A.} \bibnamefont{Kashurnikov}},
  \bibinfo{author}{\bibfnamefont{N.~V.} \bibnamefont{Prokof'ev}},
  \bibnamefont{and} \bibinfo{author}{\bibfnamefont{B.~V.}
  \bibnamefont{Svistunov}}, \bibinfo{journal}{Phys. Rev. Lett.}
  \textbf{\bibinfo{volume}{87}}, \bibinfo{pages}{120402}
  (\bibinfo{year}{2001}).

\end{thebibliography}
\end{document}